# Approximate analytical solution of the Yukawa potential with arbitrary angular momenta

M. Hamzavi[1*], M. Movahedi[2], K.-E. Thylwe[3], A. A. Rajabi[4]

[1]Department of Basic Sciences, Shahrood Branch, Islamic Azad University, Shahrood, Iran

[2]Department of Physics, Golestan University, Gorgan, Iran

[3]KTH-Mechanics, Royal Institute of Technology, S-100 44 Stockholm, Sweden

[4]Physics Department, Shahrood University of Technology, Shahrood, Iran

[*]*Corresponding author: Tel.:+98 273 3395270, fax: +98 273 3395270*

Email: majid.hamzavi@gmail.com

**Abstract**

The Yukawa potential is often used to compute bound-state normalizations and energy levels of neutral atoms. By using the generalized parametric Nikiforov-Uvarov (NU) method, we have obtained the approximate analytical solutions of the radial Schrödinger equation (SE) for the Yukawa potential. The energy eigenvalues and corresponding eigenfunctions are calculated in closed forms. Some numerical results are presented and showed that these results are in good agreement with those are obtained previously by other methods. Also, we found the energy levels of the familiar pure Coulomb potential energy levels when the screening parameter of the Yukawa potential goes to zero.

## 1. Introduction

Solutions of fundamental dynamical equations are of great interest in many fields of physics and chemistry. The exact solutions of the SE for a hydrogen atom (Coulombic) and for a harmonic oscillator represent two typical examples in quantum mechanics [1-3]. The Mie-type and pseudoharmonic potentials are also two exactly solvable potentials [4-5]. Many authors solved SE with different potentials and methods [6-17].

The Yukawa potential or static screened Coulomb potential (SSCP) [18] is given by

$$V(r) = -V_0 \frac{e^{-ar}}{r}, \tag{1}$$

where $V_0 = \alpha Z$, $\alpha = (137.037)^{-1}$ is the fine-structure constant and $Z$ is the atomic number and $a$ is the screening parameter. This potential is often used to compute bound-state normalizations and energy levels of neutral atoms [19-21] which have been studied over the past years. It is known that SSCP yields reasonable results only for the innermost states when $Z$ is large. However, for the outermost and middle atomic states, it gives rather poor results. Bound-state energies for the SSCP with $Z = 1$, have been studied in the light of the shifted large-$N$ method [22]. Chakrabarti and Das presented a perturbative solution of the Riccati equation leading to an analytic superpotential for Yukawa potential [23]. Ikhdair and Sever investigated energy levels of neutral atoms by applying an alternative perturbative scheme in solving the Schrödinger equation for the Yukawa potential model with a modified screening parameter [24]. Ikhdair and Sever also studied bound states of the Hellmann potential, which represents the superposition of the attractive Coulomb potential $-a/r$ and the Yukawa potential $b\exp(-\delta/r)$ of arbitrary strength $b$ and screening parameter $\delta$ [25].

Gonul et al. presented a new useful technique for solving the bound state problem for Yukawa-type potentials within the frame of the Riccati equation [26]. Karakoc and Boztosun applied the asymptotic iteration method to solve the radial Schrödinger equation for the Yukawa type potentials [27]. Liverts et al. used the quasi-linearization method (QLM) for solving the Schrödinger equation with a Yukawa potential [28].

This work is arranged as follows: in Section 2, the parametric generalization NU method with all the necessary formulae used in the calculations is briefly introduced. In Sec. 3 we solve SE for Yukawa potential and give energy spectra and corresponding wave functions. Some numerical results are given in this section too. Finally, the relevant conclusions and some remarks are given in section 4.

## 2. The generalized parametric of the Nikiforov-Uvarov method

To solve second order differential equations, the Nikiforov-Uvarov method can be used with an appropriate coordinate transformation $s = s(r)$ [29]

$$\psi_n''(s) + \frac{\tilde{\tau}(s)}{\sigma(s)}\psi_n'(s) + \frac{\tilde{\sigma}(s)}{\sigma^2(s)}\psi_n(s) = 0, \qquad (2)$$

where $\sigma(s)$ and $\tilde{\sigma}(s)$ are polynomials, at most of second-degree, and $\tilde{\tau}(s)$ is a first-degree polynomial. The following equation is a general form of the Schrödinger-like equation written for any potential [30]

$$\left[\frac{d^2}{ds^2} + \frac{\alpha_1 - \alpha_2 s}{s(1-\alpha_3 s)}\frac{d}{ds} + \frac{-\xi_1 s^2 + \xi_2 s - \xi_3}{[s(1-\alpha_3 s)]^2}\right]\psi_n(s) = 0. \qquad (3)$$

According to the Nikiforov-Uvarov method, the eigenfunction and eigenenergy condition become, respectively

$$\psi(s) = s^{\alpha_{12}}(1-\alpha_3 s)^{-\alpha_{12}-\frac{\alpha_{13}}{\alpha_3}} P_n^{(\alpha_{10}-1,\frac{\alpha_{11}}{\alpha_3}-\alpha_{10}-1)}(1-2\alpha_3 s), \qquad (4)$$

$$\alpha_2 n - (2n+1)\alpha_5 + (2n+1)\left(\sqrt{\alpha_9} + \alpha_3\sqrt{\alpha_8}\right) + n(n-1)\alpha_3$$
$$+ \alpha_7 + 2\alpha_3\alpha_8 + 2\sqrt{\alpha_8\alpha_9} = 0, \qquad (5)$$

where

$$\begin{aligned}
\alpha_4 &= \frac{1}{2}(1-\alpha_1), & \alpha_5 &= \frac{1}{2}(\alpha_2 - 2\alpha_3), \\
\alpha_6 &= \alpha_5^2 + \xi_1, & \alpha_7 &= 2\alpha_4\alpha_5 - \xi_2, \\
\alpha_8 &= \alpha_4^2 + \xi_3 & \alpha_9 &= \alpha_3\alpha_7 + \alpha_3^2\alpha_8 + \alpha_6.
\end{aligned} \qquad (6)$$

and

$$\begin{aligned}
\alpha_{10} &= \alpha_1 + 2\alpha_4 + 2\sqrt{\alpha_8}, & \alpha_{11} &= \alpha_2 - 2\alpha_5 + 2\left(\sqrt{\alpha_9} + \alpha_3\sqrt{\alpha_8}\right) \\
\alpha_{12} &= \alpha_4 + \sqrt{\alpha_8}, & \alpha_{13} &= \alpha_5 - \left(\sqrt{\alpha_9} + \alpha_3\sqrt{\alpha_8}\right).
\end{aligned} \qquad (7)$$

In some problems $\alpha_3 = 0$. For this type of problems when

$$\lim_{\alpha_3 \to 0} P_n^{(\alpha_{10}-1,\frac{\alpha_{11}}{\alpha_3}-\alpha_{10}-1)}(1-\alpha_3)s = L_n^{\alpha_{10}-1}(\alpha_{11}s), \qquad (8)$$

and

$$\lim_{\alpha_3 \to 0}(1-\alpha_3 s)^{-\alpha_{12}-\frac{\alpha_{13}}{\alpha_3}} = e^{\alpha_{13}s}, \qquad (9)$$

the solution given in Eq. (4) becomes [30]

$$\psi(s) = s^{\alpha_{12}} e^{\alpha_{13} s} L_n^{\alpha_{10}-1}(\alpha_{11} s). \tag{10}$$

**3. Solution of radial Schrödinger equation with the Yukawa potential**

To study any quantum physical system characterized by the empirical potential given in Eq. (1), we solve the original SE which is given in the well known textbooks [1-2]

$$\left(\frac{P^2}{2m} + V(r)\right)\psi(r,\theta,\varphi) = E\psi(r,\theta,\varphi), \tag{11}$$

where the potential $V(r)$ is taken as the Yukawa potential in Eq. (1). Using the separation method with the wave function $\psi(r,\theta,\varphi) = \frac{R(r)}{r} Y_{lm}(\theta,\varphi)$, we obtain the following radial SE as

$$\left[\frac{d^2}{dr^2} + 2m\left(E + V_0 \frac{e^{-ar}}{r}\right) - \frac{l(l+1)}{r^2}\right] R_{nl}(r) = 0. \tag{12}$$

Since the SE with the Yukawa potential has no exact solution, we use an approximation for the centrifugal term as

$$\frac{1}{r^2} \approx 4a^2 \frac{e^{-2ar}}{(1-e^{-2ar})^2}, \tag{13}$$

or equivalently

$$\frac{1}{r} \approx 2a \frac{e^{-ar}}{(1-e^{-2ar})}, \tag{14}$$

which is valid for $ar \ll 1$ [31]. Therefore, the Yukawa potential in (1) reduces to [32-34]

$$V(r) = -2aV_0 \frac{e^{-2ar}}{(1-e^{-2ar})}. \tag{15}$$

To see the accuracy of our approximation, we plotted the Yukawa potential (1) and its approximation (15) with parameters $V_0 = \sqrt{2}$, $g = \frac{a}{V_0} = 0.050$ [27], in Figure 1. Substituting (13) and (14) into (12), one obtains

$$\left[\frac{d^2}{dr^2} - \varepsilon + 4maV_0 \frac{e^{-2ar}}{1-e^{-2ar}} - l(l+1)4a^2 \frac{e^{-2ar}}{(1-e^{-2ar})^2}\right] R_{nl}(r) = 0, \tag{16}$$

where $\varepsilon = -2mE$. To solve Eq. (16) by the NU method, we use an appropriate transformation as $s = e^{-2ar}$ and Eq. (16) reduces to

$$\frac{d^2 R_{nl}(s)}{ds^2} + \frac{1-s}{s(1-s)} \frac{dR_{nl}(s)}{ds}$$

$$+ \frac{1}{s^2(1-s)^2} \left[ \frac{-\varepsilon}{4a^2}(1-s)^2 + \frac{mV_0}{a}s(1-s) - l(l+1)s \right] R_{nl}(s) = 0 \qquad (17)$$

Comparing Eq. (17) and Eq. (3), we can easily obtain the coefficients $\alpha_i$ ($i = 1, 2, 3$) and analytical expressions $\xi_j$ ($j = 1, 2, 3$) as follows

$$\alpha_1 = 1, \qquad \xi_1 = \frac{mV_0}{a} + \frac{\varepsilon}{4a^2}$$

$$\alpha_2 = 1, \qquad \xi_2 = \frac{2\varepsilon}{4a^2} + \frac{mV_0}{a} - l(l+1)$$

$$\alpha_3 = 1, \qquad \xi_3 = \frac{\varepsilon}{4a^2} \qquad (18)$$

The values of coefficients $\alpha_i$ ($i = 4, 5, ..., 13$) are found from Eqs. (6) and (7). The specific values of the coefficients $\alpha_i$ ($i = 4, 5, ..., 13$) are displayed in table 1. By using Eq. (5), we obtain the energy eigenvalues of the Yukawa potential as

$$E_{nl} = -\frac{a^2}{2m} \frac{\left( \frac{mV_0}{a} - (n+1)^2 - l(2n+l+2) \right)^2}{(n+l+1)^2} \qquad (19)$$

Somenumerical results are given in tables 2-4. In table 2, We use the parameters $\hbar = m = 1$, $V_0 = \sqrt{2}$, $g = \frac{a}{V_0} = (0.002, 0.005, 0.010, 0.020, 0.025, 0.050)$ and obtained the energy eigenvalues of the Yukawa potential for various states and compared them with others results obtained by SUSY [26], AIM [27] and numerical [35] methods. In tables 3 and 4, we show the numerical results with parameter set $\hbar = 2m = 1$, $a = 0.2 fm^{-1}$ and compared with those of other methods.

When the screening parameter $a$ approaches zero, the potential (1) reduces to a Coulomb potential. Thus, in this limit the energy eigenvalues of (19) become the energy levels of the pure Coulomb interaction, i.e.

$$E_{n,l_{Coulomb}} = -\frac{1}{2} m \frac{V_0^2}{n'^2} \qquad (20)$$

where $n' = n + l + 1$ [1,2].

To find corresponding wave functions, referring to table 1 and Eq. (8), we find the radial wave functions as

$$R_{nl}(s) = s^{\sqrt{\frac{\varepsilon}{4a^2}}} (1-s)^{l+1} P_n^{(2\sqrt{\frac{\varepsilon}{4a^2}}, 2l+1-\sqrt{\frac{\varepsilon}{4a^2}})} (1-2s), \qquad (21)$$

or, by substituting $s = e^{-2ar}$;

$$R_{nl}(s) = N e^{-\sqrt{\varepsilon}r} (1-e^{-2ar})^{l+1} P_n^{(2\sqrt{\frac{\varepsilon}{4a^2}}, 2l+1-\sqrt{\frac{\varepsilon}{4a^2}})} (1-2e^{-2ar}), \qquad (22)$$

where $N$ is normalization constant.

## 4. Conclusions and Remarks

In this article, we have obtained the bound state solutions of the Schrödinger equation for the Yukawa potential by using the parametric generalization of the Nikiforov-Uvarov method. The energy eigenvalues and corresponding eigenfunctions are obtained by this method. Some numerical results are given in table 2 and the comparison of calculation results with the accurate numerical values has been proven the success of the formalism. It is found that when the screening parameter $a$ goes to zero, the energy levels approach to the familiar pure Coulomb potential energy levels. The aim of solving the Yukawa potential returns to the following reasons: First, in the low screening region where the screening parameter $a$ is small (i.e., $a \ll 1$), the potential reduces to the Killingbeck potential [36-38], i.e., $V(r) = ar^2 + br - c/r$, where $a$, $b$ and $c$ are potential constants that can be obtained after making expansion to the Yukawa potential. Second, it can also be reduced into the Cornell potential [39, 40], i.e., $V(r) = br - c/r$. These two potentials are usually used in the study of mesons and baryons. Third, when the screening parameter approaches to zero, the Yukawa potential turns to become the Coulomb potential.

**Acknowledgments**

We would like to thank the kind referees for positive suggestions which have improved the present work.

**Table 1.** The specific values for the parametric constants necessary for the energy eigenvalues and eigenfunctions

| constant | Analytic value |
|---|---|
| $\alpha_4$ | $0$ |
| $\alpha_5$ | $-\dfrac{1}{2}$ |
| $\alpha_6$ | $\dfrac{1}{4} + \dfrac{mV_0}{a} + \dfrac{\varepsilon}{4a^2}$ |
| $\alpha_7$ | $-\dfrac{2\varepsilon}{4a^2} - \dfrac{mV_0}{a} + l(l+1)$ |
| $\alpha_8$ | $\dfrac{\varepsilon}{4a^2}$ |
| $\alpha_9$ | $\left(l+\dfrac{1}{2}\right)^2$ |
| $\alpha_{10}$ | $1 + 2\sqrt{\dfrac{\varepsilon}{4a^2}}$ |
| $\alpha_{11}$ | $2l + 3 + \sqrt{\dfrac{\varepsilon}{4a^2}}$ |
| $\alpha_{12}$ | $\sqrt{\dfrac{\varepsilon}{4a^2}}$ |
| $\alpha_{13}$ | $-\left(l+1+\sqrt{\dfrac{\varepsilon}{4a^2}}\right)$ |

**Table 1.** The energy eigenvalues (in $fm^{-1}$) of the Yukawa potential in units $\hbar = m = 1$. We set $V_0 = \sqrt{2}$ and $a = gV_0$ for comparison with other methods.

| State | g | Approximation (Present Calculations) | AIM [27] | SUSY [26] | Numerical [35] |
|---|---|---|---|---|---|
| 1s | 0.002 | −0.99600 | −0.99600 | −0.99601 | −0.99600 |
|    | 0.005 | −0.99002 | −0.99003 | −0.99004 | −0.99000 |
|    | 0.010 | −0.98010 | −0.98014 | −0.98015 | −0.98010 |
|    | 0.020 | −0.96040 | −0.96059 | −0.96059 | −0.96060 |
|    | 0.025 | −0.95062 | −0.95092 | −0.95092 | −0.95090 |
|    | 0.050 | −0.90250 | −0.90363 | −0.90363 | −0.90360 |
| 2s | 0.002 | −0.24601 | −0.24602 | −0.24602 | −0.24600 |
|    | 0.005 | −0.24010 | −0.24014 | −0.24015 | −0.24010 |
|    | 0.010 | −0.23040 | −0.23058 | −0.23059 | −0.23060 |
|    | 0.020 | −0.21160 | −0.21229 | 0.21230 | −0.21230 |
|    | 0.025 | −0.20250 | −0.20355 | −0.20355 | −0.20360 |
|    | 0.050 | −0.16000 | −0.16354 | −0.16351 | −0.16350 |
| 2p | 0.002 | −0.24601 | −0.24601 | −0.24602 | −0.24600 |
|    | 0.005 | −0.24010 | −0.24012 | −0.24012 | −0.24010 |
|    | 0.010 | −0.23040 | −0.23049 | −0.23049 | −0.23050 |
|    | 0.020 | −0.21160 | −0.21192 | −0.21192 | −0.21190 |
|    | 0.025 | −0.20250 | −0.20298 | −0.20299 | −0.20300 |
|    | 0.050 | −0.16000 | −0.16148 | −0.16144 | −0.16150 |
| 3p | 0.002 | −0.10714 | −0.10716 | −0.10716 | −0.10720 |
|    | 0.005 | −0.10133 | −0.10141 | −0.10142 | −0.10140 |
|    | 0.010 | −0.09201 | −0.09230 | −0.09231 | −0.09231 |
|    | 0.020 | −0.07471 | −0.07570 | −0.07570 | −0.07570 |
|    | 0.025 | −0.06673 | −0.06815 | −0.06814 | −0.06816 |
|    | 0.050 | −0.03361 | −0.03711 | −0.03739 | −0.03712 |
| 3d | 0.002 | −0.10714 | −0.10715 | −0.10715 | −0.10720 |
|    | 0.005 | −0.10133 | −0.10136 | −0.1014 | −0.10140 |
|    | 0.010 | −0.09201 | −0.09212 | −0.09212 | −0.09212 |
|    | 0.020 | −0.07471 | −0.07503 | −0.07502 | −0.07503 |
|    | 0.025 | −0.06673 | −0.06714 | −0.06713 | −0.06715 |
|    | 0.050 | −0.03361 | −0.03383 | −0.03388 | −0.03383 |

**Table 2.** The same as in Table 1, but $\hbar = 2m = 1$, $a = 0.2\,fm^{-1}$ and $n = 0$.

| $V_0$ | $l$ | Approximation (Present Calculations) | AIM [27] | SUSY [26] | Numerical [23] | Analytical [23] |
|---|---|---|---|---|---|---|
| 4 | 0 | −3.2400 | −3.2564 | −3.2563 | −3.2565 | −3.2199 |
| 8 | 0 | −14.4400 | −14.4581 | −14.4581 | −14.4571 | −14.4199 |
|   | 1 | −2.5600 | −2.5836 | −2.5830 | −2.5836 | −2.4332 |
| 16 | 0 | −60.8400 | −60.8590 | −60.8590 | −60.8590 | −60.8193 |
|    | 1 | −12.9600 | −12.9910 | −12.9908 | −12.9910 | −12.8375 |
| 24 | 0 | −139.2400 | −139.2593 | −139.2590 | −139.2594 | −139.2201 |
|    | 1 | −31.3600 | −31.39381 | −31.3937 | −31.3938 | −31.2385 |
|    | 2 | −11.5600 | −11.5959 | −11.5951 | −11.5959 | −11.2456 |

**Table 3.** The same as in Table 2, but $n > 0$.

| $V_0$ | $n$ | $l$ | Approximation (Present Calculations) | AIM [27] | SUSY [26] | Numerical [23] | Analytical [23] |
|---|---|---|---|---|---|---|---|
| 16 | 1 | 0 | −12.9600 | −13.0273 | −13.0270 | −13.0273 | −13.0326 |
|    | 2 | 0 | −4.2711 | −4.3941 | −4.3937 | −4.3720 | −4.4057 |
|    | 1 | 1 | −4.2711 | −4.3621 | −4.3612 | −4.3480 | −4.3886 |
| 24 | 1 | 0 | −31.3600 | −31.4312 | −31.431 | −31.4356 | −31.4313 |
|    | 2 | 0 | −11.5600 | −11.6998 | −11.6990 | −1.6998 | −11.7093 |
|    | 3 | 0 | −4.8400 | −5.0441 | −5.0448 | −5.0442 | −5.0590 |
|    | 4 | 0 | −1.9600 | −2.2033 | −2.2194 | −2.2033 | −2.2237 |
|    | 1 | 1 | −11.5600 | −11.6652 | −11.664 | −11.6653 | −11.6839 |
|    | 2 | 1 | −4.8400 | −5.0134 | −5.0133 | −5.0135 | −5.0541 |
|    | 3 | 1 | −1.9600 | −2.1770 | −2.1908 | −2.1770 | −2.2414 |
|    | 1 | 2 | −4.8400 | −4.9515 | −4.9504 | −4.9516 | −5.0085 |
|    | 2 | 2 | −1.9600 | −2.1241 | −2.1337 | −2.1241 | −2.2428 |

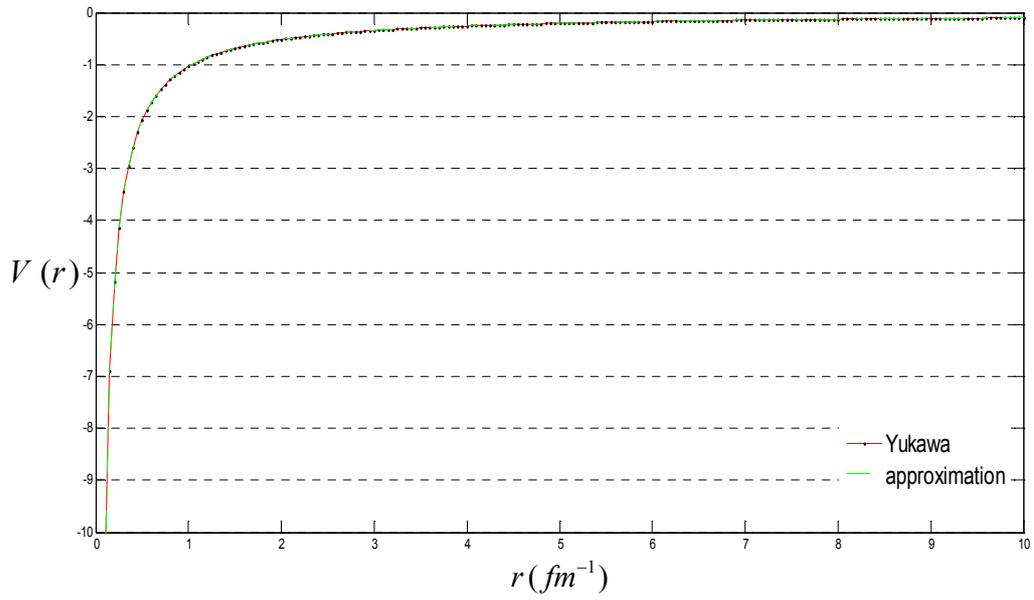

**Figure 1:** The Yukawa potential (red line) and its (red line) and its approximation in Eq. (15) (green line).